\documentclass[12pt]{article}
\usepackage{epsfig} 
\def\eqn#1{eq.~(\ref{#1})}

\newcommand{\be}{\begin{equation}} 
\newcommand{\ee}{\end{equation}} 
 
\newcommand{\as}{\alpha_s}

\begin{document}  

\begin{titlepage}
\begin{flushright}
IPPP/02/65 \\
DCPT/02/130
\end{flushright}
\vspace{0.5cm}
\begin{center}
{\Large{\bf Gluon induced contributions to }} \\
\vskip 0.2cm
{\Large{\bf $Z\gamma$ production at hadron colliders} }
 
\vskip 1.2cm
{\bf K. L. Adamson$^{a}$, 
D. de Florian$^{b}$
and  A. Signer$^a$}
\vskip .8cm
\vskip .3cm
{\it $^a$ IPPP, Department of Physics, University of Durham,
Durham DH1 3LE, England}
\vskip .3cm
{\it $^b$ Departamento de F\'\i sica, Universidad de Buenos Aires, Argentina}
\vskip 4.cm

\vspace*{0.3cm}  

\hspace{1cm}
 
\large
{\bf Abstract} \\
\end{center}

We study the contribution of gluon induced partonic subprocesses to
$Z\gamma$ pair production at hadron colliders. These processes
contribute only at next-to-next-to-leading order but are potentially
enhanced by two factors of the gluon parton densities. However, we
find that their contribution is modest and that next-to-leading order
calculations give reliable predictions.

\vspace{0.5cm}
\normalsize
   \end{titlepage}
\newpage

\setcounter{page}{1}  
\pagestyle{plain}


\section{Introduction}  

The substantially increased number of $Z\gamma$ pairs that are
expected to be produced in Run II at the Tevatron will allow an
improvement to previous studies~\cite{Abe:1994zj, Abachi:1997ef,
Abbott:1998hr} of $Z\gamma$-pair production at hadron
colliders. Accordingly, the $ZZ\gamma$ and $Z\gamma\gamma$ couplings,
which are absent in the Standard Model, can be investigated in more
detail. Assuming no deviations from the Standard Model are found,
tighter limits on such anomalous couplings will be possible.

In order to fully exploit the experimental data it is important to
have theoretical predictions of sufficient accuracy. Next-to-leading
order cross sections for this process have been calculated quite some
time ago \cite{Ohnemus:1992jn}. Subsequently, the theoretical
predictions have been improved by taking into account the leptonic
decay of the $Z$-boson, anomalous couplings and full leptonic
correlations \cite{Ohnemus:qp, Baur:1997kz, Dixon:1998py,
DeFlorian:2000sg}.

For some distributions, the next-to-leading order corrections have
been found to be very large. In particular in the large transverse
momentum region, which is interesting for investigating anomalous
couplings, the corrections can increase the theoretical prediction of
the cross section by several 100\%. The reason for this large
corrections is that at NLO new partonic processes contribute. Whereas
the tree-level calculation only incorporates the partonic process
$q\bar{q} \to Z \gamma \to \ell \bar{\ell} \gamma$, at next-to-leading
order also processes with a gluon in the initial state such as $q g
\to Z \gamma q \to \ell \bar{\ell} q \gamma$ contribute. The large
gluon density at low $x$ overcomes the suppression by an additional
factor of $\as$. Thus, even though this contribution is formally
next-to-leading order, numerically it can be as important as the
leading-order term.

This immediately leads to the question whether at NNLO a similar
effect occurs. Higher-order corrections to partonic processes that are
already present at NLO are not expected to lead to large corrections.
However, at NNLO a new class of partonic processes have to be taken
into account, namely processes with two gluons in the initial state.
Such processes are suppressed by two factors of $\as$ but are
potentially enhanced by two factors of the gluon parton distribution
function. If there is a similar effect as at NLO, this could lead to
corrections as important as the Born term.

This question has been studied for $WZ$ and $W\gamma$ production
\cite{Adamson:2002jb, Adamson:thesis}. For these processes it has been
found that the gluon induced NNLO corrections are negligible. However,
there is an important difference in $Z\gamma$ production with respect
to $WZ,\ W\gamma$ production. In fact, there are generally two classes
of contributions for vector-boson pair production with two gluons in
the initial state. There are tree-level processes $gg\to VVq\bar{q}$
and loop processes $gg\to VV$, where $V$ denotes some vector boson. In
the case of $WZ$ and $W\gamma$ the loop processes vanish due to charge
conservation but for $Z\gamma$ this is not the case. The study in
\cite{Adamson:2002jb, Adamson:thesis} showed that the tree-level
processes result in a very small contribution for $WZ$ and
$W\gamma$. While it is natural to expect that this is also the case
for $Z\gamma$ production, for this process there remains the question
of how important the gluon induced loop corrections are.

The loop induced process $gg\to Z\gamma$ has beeen studied previously
\cite{Ametller:1985di, vanderBij:1988fb, Glover:1988fe}, where it has
been found that these contributions can be quite large but are not
dominant.

This work extends the previous analyses in that both classes of gluon
induced corrections to $Z\gamma$ production are included and anomalous
$ZZ\gamma$ and $Z\gamma\gamma$ couplings as well as full leptonic
correlations are taken into account. Even though the gluon induced
correction is substantially more important for $Z\gamma$ production
than for $W\gamma,\ WZ$ it can still be safely neglected. In fact, not
only are the contributions from tree-level processes tiny (as in the
$W\gamma,\ WZ$ case) but also the loop processes are smaller than
anticipated. The latter effect is entirely due to a change in the
parton distribution function used in our analysis compared to
\cite{vanderBij:1988fb, Glover:1988fe}.

\section{Calculation}  

Our aim is to compute the part of the NNLO corrections to $Z\gamma$
pair production that is enhanced by two factors of the gluon parton
distribution function. This consists of two parts. Firstly, we have to
compute the tree-level amplitudes $gg\to q\bar{q}Z\gamma \to
q\bar{q}\ell\bar{\ell}\gamma$ and integrate over the quark-antiquark
final state phase space. Secondly, we have to evaluate the loop
diagrams $gg\to Z\gamma \to \ell\bar{\ell}\gamma$.

The contributions of the tree-level diagrams $gg\to q\bar{q}Z\gamma
\to q\bar{q}\ell\bar{\ell}\gamma$ can be computed in a very similar
manner as for $gg\to q\bar{q}W\gamma$, presented in
\cite{Adamson:2002jb}. This is a double bremsstrahlung NNLO
contribution and in general is difficult to integrate over the phase
space. However, due to the simple structure of the soft and collinear
limits of the corresponding matrix elements, the integration over the
singular regions of the phase space is much simpler than for a general
NNLO calculation. In fact, as discussed in~\cite{Adamson:2002jb}, the
only singularities we are concerned with are single and double initial
state collinear singularities. They come from the region of phase
space where the two incoming gluons independently split into a
quark-antiquark pair.

In order to perform the integration over phase space we use a
generalization of the subtraction method presented
in~\cite{Frixione:1995ms}. The singularities that arise upon
integration over the region of phase space with the final state
(anti)quark collinear to the incoming gluons are absorbed into the
parton distribution functions. It should be noted that this
contribution is factorization scheme dependent. Since we use parton
densities obtained in the $\overline{MS}$-scheme we evaluate the
partonic cross section in this scheme. We should also mention that the
factorization scheme dependence is beyond the approximation we make
since it is not enhanced by two factors of the gluon density.

The tree-level diagrams for $gg\to q\bar{q} V V$ have been computed
previously~\cite{Baur:1990mr}. We included anomalous couplings
$ZZ\gamma$ and $Z\gamma\gamma$. They are incorporated by using the
$Z_\alpha(q_1)\gamma_\beta(q_2)Z_\mu(p)$ vertex
\cite{Gounaris:1999kf,Hagiwara:1986vm} 
\begin{eqnarray}
 \label{eq:zzy-vertex}
\Gamma_{Z\gamma Z}^{\alpha\beta\mu}(q_1,q_2,p)&=&
\frac{i(p^2-q_1^2)}{M_Z^2} \left(
 h_1^Z(q_2^\mu \, g^{\alpha\beta}-q_2^\alpha \, g^{\mu\beta}) +
\frac{h_2^Z}{M_Z^2} 
     p^\alpha(p \cdot  q_2 \, g^{\mu\beta}-q_2^\mu \,p^\beta ) 
\right.   \nonumber \\
&& \left. 
-\, h_3^Z \varepsilon^{\mu\alpha\beta\nu} q_{2\nu} -
\frac{h_4^Z}{M_Z^2}
\varepsilon^{\mu\beta\nu\sigma}p^\alpha p_\nu \, q_{2\sigma} \right)
\end{eqnarray}
where it should be noted that the $Z$-boson with momentum $p$ is
off-shell and the other two gauge bosons are on-shell. The vertex for
$Z_\alpha(q_1)\gamma_\beta(q_2)\gamma_\mu(p)$ is as in
\eqn{eq:zzy-vertex}, but with $q_1^2 \rightarrow 0$ and  
$h_i^Z \rightarrow h_i^\gamma$.

The inclusion of anomalous couplings also necessitates the recalculation
of the gluon induced loop diagrams. First of all, there are box
diagrams. They have been calculated some time
ago~\cite{Ametller:1985di, vanderBij:1988fb, Glover:1988fe} but these
results were obtained summing over helicities and did not include the
decay of the $Z$-boson. However, since the axial coupling of the $Z$
does not contribute, the amplitude can also be extracted from (the
$N_F$-part of) the QCD amplitudes $gg\to q\bar{q}\gamma$
\cite{Bern:1994fz, Signer:1995nk}. In fact we simply have to change
the gluon that decays into the $q\bar{q}$ pair into a $Z$. This can be
done by picking the relevant subamplitudes of the $gg\to
q\bar{q}\gamma$ process and changing the couplings accordingly.

In addition to the box diagrams there are triangle diagrams with a
triple $ZZ\gamma$ or $Z\gamma\gamma$ vertex. These diagrams are only
present for anomalous couplings. Furthermore, due to Yang's
theorem~\cite{Yang:rg} this contribution is only non-vanishing because
of off-shell effects. It should also be stressed, that the vertex of
\eqn{eq:zzy-vertex} does not include all possible anomalous
couplings. The vertex in \eqn{eq:zzy-vertex} does not include terms
$\sim p^\mu$ since it is assumed that the gauge bosons couple to a
conserved current. However, in our situation this is not the case and
aditional anomalous couplings are possible \cite{Gounaris:2000dn}.
There is one additional CP conserving ($p^{\mu} q_1^\sigma q_2^\nu
\varepsilon_{\sigma\nu\alpha\beta}$) and two additional CP
violating couplings ($p_{\mu} g_{\alpha\beta}$ and $p_{\mu}
(p-q_2)_{\alpha} (p-q_1)_{\beta}$). In our calculation we have not
implemented these additional anomalous couplings, since --- given the
small size of the gluon induced corrections --- it is very unlikely
that useful information about these couplings can be extracted.

The effects of the top quark are taken into account by performing an
expansion in $1/m_t$ and neglecting terms of order $(1/m_t)^4$. This
approach is motivated by assuming that the partonic center of mass
energy $\sqrt{\hat{s}}< m_t$. Even though this is not generally true
at the LHC, the fact that the gluon distribution is strongly peaked
towards small values of $x$, makes this a reasonably good
approximation. 

To deal with the problem of photon isolation, we will use the procedure 
introduced by Frixione~\cite{Frixione:1998jh}, i.e. we reject  all events
unless the transverse hadronic momentum deposited in a cone of size
$R_0$ around the momentum of the photon fulfills the following
condition
\begin{equation} 
\label{eq:isolation} 
\sum_i p_{Ti}\, \theta(R -R_{i\gamma}) \le p_{T\gamma}  
\left( \frac{1-\cos R}{1-\cos R_0}\right) \,, 
\end{equation} 
for all $R \le R_0$, where the `distance' in pseudorapidity and azimuthal 
 angle is defined by $R_{i\gamma}=\sqrt{(\eta_i-\eta_\gamma)^2 +  
(\phi_i-\phi_\gamma)^2 }$.

\section{Numerical Results}  
  
For the numerical results presented in this section we follow closely
the analysis presented in~\cite{Adamson:2002jb}. We use the MRST~2001
parton distribution functions~\cite{Martin:2001es} with the one-loop
expression for the coupling constant ($\alpha_s(M_Z) = 0.119$). 
Strictly speaking, for the $gg$ induced processes we should
use a NNLO parton distribution. However, it is not expected that a
change to the not yet fully available NNLO distributions would alter
our conclusions.  The factorization and renormalization scales are
fixed to $\mu_F=\mu_R= \sqrt{M_Z^2 + (p_T^\gamma)^2}$.

The mass of the $Z$ has been set to $M_Z=91.187$~GeV. For the
couplings of the vector bosons with the quarks we use $\alpha =
\alpha(M_Z)= 1/128$ whereas for the photon coupling we use $\alpha =
1/137$. Note that we do not include the branching ratios for the decay
of the $Z$ into leptons.

For our numerical results we use a set of standard cuts. Charged
leptons are required to have $p_T>20$~GeV and $\eta < 2.5$. The photon
transverse momentum cut we use is $p_T^\gamma > 20$~GeV, while for the
isolation prescription in \eqn{eq:isolation} we set $R_0=1$.

\begin{figure}[htb]
\begin{center}
\begin{tabular}{c}
\epsfxsize=12truecm
\put(70,210){${\rm d}\sigma /{\rm d} p_T^\gamma\ $[pb/GeV]}
\epsffile{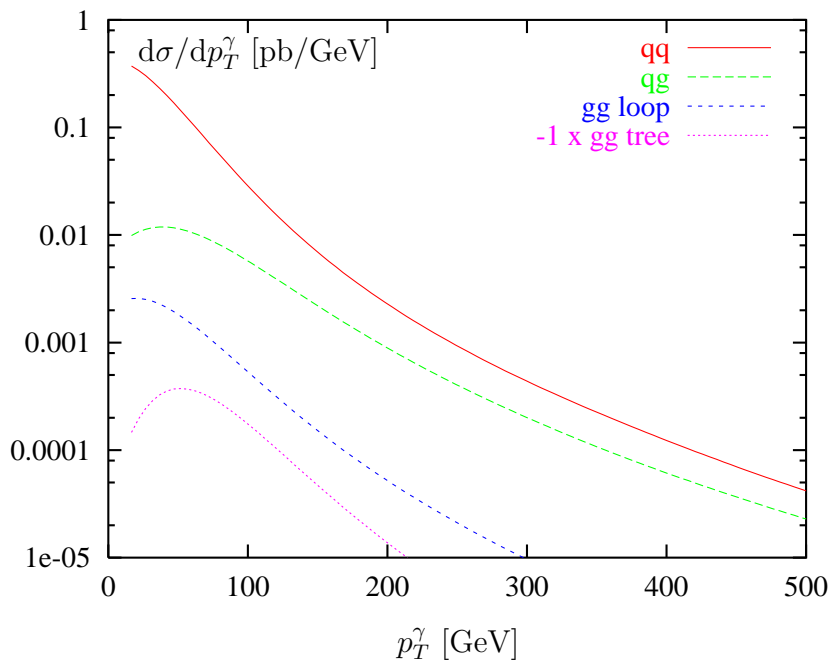}\\
\put(168,65){$p_T^\gamma\ $[GeV]}
\vspace{-2.9cm}
\end{tabular}
\end{center}
\caption[$gg$ induced part of $Z\gamma$]
{\it The contribution to the $p_T^\gamma$ distribution in
  $Z\gamma$ production at the LHC separated into the various partonic
  initial states: $q\bar{q}$ (solid line), $g q$ (dashed) and
  $gg$. The gluon induced contribution is separated into the loop and
  bremsstrahlung part.} 
\label{fig:zyboth}
\end{figure}

Since the potential enhancement due to the gluon densities is larger
for increasing center of mass energy, we restrict ourselves to
studying the situation at the LHC. Thus, we consider $Z\gamma$
production in proton--proton collision at an energy $\sqrt{s} =
14~$TeV. 
 
The canonical quantity we use to investigate the importance of the
gluon induced corrections is the transverse momentum of the photon. In
Figure~\ref{fig:zyboth} we show the contribution of the various
initial partonic states to ${\rm d}\sigma/{\rm d} p_T^\gamma$. The
solid line shows the contribution at NLO of the initial state
$q\bar{q}$. This is the dominant part and differs only slightly from
the tree-level result. The $g q$ (and $g\bar{q}$) initial states enter
at NLO and can result in a correction as big as 70\% which is shown as
the dashed line. Finally, there are the $gg$ induced processes. We
split the result into the loop part (short dashed) and the tree-level
bremsstrahlung part (dotted). The latter is negative and very
small. This is very similar to what has been found for $WZ$ and
$W\gamma$ production~\cite{Adamson:2002jb}. The loop part is
substantially bigger but still does not exceed 5\%. Note that differs
quite a bit from what has been found in previous
analyses~\cite{vanderBij:1988fb, Glover:1988fe}. We checked that this
is entirely due to an update in the gluon distribution function. In
fact, our results agree with those in ~\cite{vanderBij:1988fb,
Glover:1988fe} if we use the same parton distribution functions.

The situation does not change if anomalous couplings are added. The
relative importance of the gluon induced processes is still
small. Thus, there is no hope of gaining any information about
non-standard anomalous couplings that in principle contribute to this
process.

We also investigated other quantities and the picture remains the
same. The loop corrections have more or less the expected size but are
never very important.  The reason for the smallness of the
bremsstrahlung corrections is the following: in the important region
of small partonic center of mass energy $\sqrt{\hat{s}}$,
corresponding to small $x$, the partonic cross section ${\rm
d}\hat{\sigma}$ is small and turns negative. Only for increasing
$\sqrt{\hat{s}}$ we find that ${\rm d}\hat{\sigma}$ is of the expected
size, i.e. suppressed by two orders of $\as$. However, in this region
$x$ is not small anymore and, thus, there is no enhancement due to the
gluon distribution. This might well be related to the very simple
structure of the singularities of the partonic cross section. For a
process with a more complicated structure, e.g. involving $t$-channel
gluon exchange, the bremsstrahlung corrections could easily be
significantly higher.

\section{Conclusions}  

We studied the contribution of the partonic processes $gg\to q
\bar{q}Z\gamma$ and $gg\to Z\gamma$ to the production of $Z\gamma$
pairs at hadron colliders. These contributions are enhanced by two
factors of the large gluon density. This could potentially overcome
the ${\cal O}(\as^2)$ suppression. However, we found that under
no circumstances are these contributions particularly important. In
fact, they are even substantially smaller than anticipated from
previous analyses~\cite{vanderBij:1988fb, Glover:1988fe}, a change
that is due to using updated parton distribution functions.

In summary, the next-to-leading order cross section for $Z\gamma$
production provide us with a reasonably precise theoretical
prediction. The large NLO corrections due to the opening of a new
partonic channel with $qg$ initial states are taken into
account. Higher order corrections to these partonic processes are
expected to be well under control and the new partonic channels that
open at NNLO, namely $gg$ initial state processes, do not result in
large corrections.

\subsection*{Acknowledgments}


It is a pleasure to acknowledge very useful discussions with L.~Dixon.
 KLA acknowledges support from a PPARC
studentship. DdF is supported by Conicet and Fundaci\'on Antorchas.
  


\end{document}